%% file: main.tex
\documentclass[%
 aip,jcp,
 amsmath,amssymb,
 reprint,
]{revtex4-2}

\usepackage{graphicx}% Include figure files
\usepackage{dcolumn}% Align table columns on decimal point
\usepackage{bm}% bold math
\usepackage{caption}
\usepackage{subcaption}

\usepackage[utf8]{inputenc}
\usepackage[T1]{fontenc}
\usepackage{mathptmx}
\usepackage{listings}
\usepackage{rotating} % Rotating table

\usepackage{mathtools}
\usepackage{color}
\usepackage{dcolumn} % decimal align in tables
\usepackage{bm} % bold math
\usepackage{graphicx}
\usepackage{multirow} % for table cells to span rows
\usepackage{pifont} % for checkmarks
\usepackage{epsfig}
\usepackage{amsmath} % matrix
\usepackage{float}
\usepackage{booktabs}
\usepackage{tabularx}
\usepackage{natbib}
\usepackage{gensymb}
\usepackage{rotating}
\usepackage{threeparttable}
\usepackage{comment}
\usepackage[normalem]{ulem}
\usepackage{xr-hyper}
\usepackage[hidelinks]{hyperref} % allows hyperlinking for references

\usepackage{xcolor}

\usepackage{xr}
\makeatletter

\newcommand*{\addFileDependency}[1]{% argument=file name and extension
\typeout{(#1)}% latexmk will find this if $recorder=0
\@addtofilelist{#1}
\IfFileExists{#1}{}{\typeout{No file #1.}}
}\makeatother

\newcommand*{\myexternaldocument}[1]{%
\externaldocument{#1}%
\addFileDependency{#1.tex}%
\addFileDependency{#1.aux}%
}
\myexternaldocument{si}

\begin{document}

\author{Grier M. Jones}
\affiliation{
             Department of Chemistry,
             University of Tennessee,
             Knoxville, TN 37996}
  
\author{Run. R. Li}
\affiliation{
             Department of Chemistry and Biochemistry,
             Florida State University,
             Tallahassee, FL 32306-4390}          
             
\author{A. Eugene DePrince III}
\email{adeprince@fsu.edu}
\affiliation{
             Department of Chemistry and Biochemistry,
             Florida State University,
             Tallahassee, FL 32306-4390}

\author{Konstantinos D. Vogiatzis}
\email{kvogiatz@utk.edu}
\affiliation{
             Department of Chemistry,
             University of Tennessee,
             Knoxville, TN 37996}

\title{
%Bridging the Gap between Two-electron Reduced-density-matrix Theory and FCI with Data-driven Methods}
Data-driven Refinement of Electronic Energies from Two-Electron Reduced-Density-Matrix Theory}

\begin{abstract}

The exponential computational cost of describing strongly correlated electrons can be mitigated by adopting a reduced-density-matrix (RDM)-based description of the electronic structure. While variational two-electron RDM (v2RDM) methods can enable large-scale calculations on such systems, the quality of the solution is limited by the fact that only a subset of known necessary $N$-representability constraints can be applied to the 2RDM in practical calculations. Here, we demonstrate that violations of partial three-particle (T1 and T2) $N$-representability conditions, which can be evaluated with knowledge of only the 2RDM, can serve as \textit{physics-based} features in a machine-learning (ML) protocol for improving energies from v2RDM calculations that consider only two-particle (PQG) conditions. Proof-of-principle calculations demonstrate that the model yields substantially improved energies, relative to reference values from configuration-interaction-based calculations.

\end{abstract}

\maketitle

%\section{Introduction}
%\input{introduction}

%\label{SEC:INTRODUCTION}

%\noteg{If we submit to JCPL we will have to remove the section headings}

An active space of strongly correlated electrons is challenging to describe computationally because the complexity of the exact configuration interaction (CI) wave function grows exponentially with the number of active electrons and orbitals. As a result, the largest reported complete active space self-consistent field (CASSCF)\cite{Roos:1980:157,Siegbahn:1980:323,Siegbahn:1981:2384,Roos:1987:399} calculation to date involved only 22 electrons distributed among 22 orbitals [a (22e, 22o) active space].\cite{Vogiatzis:2017:184111} 
In light of these difficulties, a large body of work has attempted to circumvent the exponential scaling of exact CI through various approximate CI schemes,\cite{Olsen:1988:2185,Malmqvist:1990:5477,Fleig:2001:4775,Ma:2011:044128,Manni:2013:3375,Thomas:2015:5316,Manni:2016:1245,HeadGordon20_2340,Evangelista16_161106} as well as via alternative representations of the electronic structure, including the density matrix renormalization group (DMRG) approach\cite{Ghosh:2008:144117,Yanai:2009:2178,Wouters:2014:1501,Sun:2017:291,Ma:2017:2533} and two-electron reduced density matrix (2RDM) methods.\cite{Mazziotti08_134108,DePrince16_2260}

%2RDM theory seems ostensibly to be a desirable alternative to CI-based descriptions of correlated electronic structure. 

The energy of a many-electron system can be represented exactly in terms of the 2RDM, and the 2RDM itself is a compact mathematical object (relative to the complexity of the wave function). Hence, a 2RDM-based representation of electronic structure seems natural, ostensibly making 2RDM theory a desirable alternative to CI. Unfortunately, a variety of issues plague 2RDM-based calculations,\cite{Ayers09_5558,Bultinck10_114113,Cooper11_054115,Ayers12_014110,DePrince19_032509,DePrince22_5966} and these issues are rooted in a singular challenge: without explicit knowledge of a wave function that maps to the 2RDM, a large number of non-trivial constraints must be applied to the 2RDM to guarantee that it is derivable from an $N$-electron density matrix or an ensemble of such density matrices. Such a 2RDM is said to be ``$N$-representable.''\cite{Coleman63_668} %(specifically, the 2RDM for these cases would be pure-state or ensemble-state $N$-representable, respectively). 

The ensemble $N$-representability problem is, in principle, solved, at least in the sense that a complete hierarchy of constraints on the 2RDM has been proposed.\cite{Mazziotti12_263002, Mazziotti12_062507, Mazziotti23_153001} In practical variational 2RDM (v2RDM) calculations,\cite{Husimi:1940:264, Lowdin:1955:1474,Mayer:1955:1579,Percus64_1756,Rosina75_868,Rosina75_221,Garrod75_300,Rosina79_1366,Erdahl79_147,Fujisawa01_8282,Mazziotti02_062511,Mazziotti06_032501,Percus04_2095,Zhao07_553,Lewin06_064101,Bultinck09_032508,DePrince16_423,DeBaerdemacker11_1235,Mazziotti11_5632, VanNeck15_4064,DeBaerdemacker18_024105,Mazziotti17_084101,Ayers09_5558,Bultinck10_114113,Cooper11_054115,DePrince19_032509,Mazziotti08_134108,DePrince16_2260,Mazziotti16_153001, DePrince19_6164} however, one imposes a only subset of necessary $N$-representability conditions;
%, truncated at the $p$-particle level (with $p \geq N$);\cite{Erdahl01_042113} 
such conditions usually include the two-particle (PQG) conditions\cite{Percus64_1756} or the partial three-particle conditions known as T1 and T2.\cite{Erdahl78_697,Percus04_2095} At the PQG level, in particular, large numbers of strongly correlated electrons can be efficiently treated; for example, a v2RDM-based calculation involving a (64e, 64o) active space can be completed in a matter of hours.\cite{DePrince19_6164} That said, the quantitative accuracy of PQG-level calculations can be poor, with the correlation energy often overestimated by as much as 20\%, even in small systems near their equilibrium geometries.\cite{Mazziotti_11_052506} Hence, it is clear that partial three-particle conditions ({e.g., T2) or even full three-particle conditions (three-positivity or 3POS)\cite{Mazziotti06_032501,DePrince21_174110} may be necessary to achieve quantitative accuracy. The challenge is that imposing these additional conditions comes at a much higher floating-point cost. 

Machine learning (ML) is  an emerging tool in electronic structure theory for recovering the electronic energies,\cite{ramakrishnan2015big,schutt2018schnet,Welborn2018, Cheng2019, Cheng2019reg, townsend2020transferable,Husch2021,jones2023data}  wave functions,\cite{townsend2019data,hermann2020deep}  and molecular properties of post-Hartree-Fock methods.\cite{Dral2023,chen2023benchmark}
Specifically, for ML applications to wave function methods, representative examples include coupled-cluster with singles-and-doubles excitations (CCSD) theory, where ML is used to learn the two-electron amplitudes,\cite{townsend2019data,townsend2020transferable} and the CI problem where ML allows the efficient selection of important configurations.\cite{coe2018machine,coe2019machine,yang2020artificial,pineda2021chembot,jeong2021active}
%These include methods related to the CI problem, such as using ML for the efficient selection of configuration coefficients.\cite{coe2018machine,coe2019machine,yang2020artificial,pineda2021chembot,jeong2021active}
With respect to RDMs, a so-called density tensor representation has been proposed\cite{peyton2020machine} to predict CCSD electronic energies and dipole moments from the one-electron reduced density matrix (1RDM), and, more recently, Sager-Smith and Mazziotti\cite{sager2022reducing} have demonstrated that geminal occupations (the eigenvalues of the 2RDM) can be estimated from correlation temperatures extracted from a convolutional neural network.  
%With respect to RDMs, the density tensor representation can be used to learn electronic energies and dipole moments\cite{peyton2020machine} and the recent publication of Sager-Smith and Mazziotti\cite{sager2022reducing} shows that 2RDM-based coupled-cluster singles and doubles (CCSD) and CASSCF energies can be recovered via the prediction of correlation temperatures, which are used to define Boltzmann-like distributions of geminal populations, from geminal energies and Hartree-Fock correlation temperatures using convolutional neural networks.

The goal of the present work is the prediction of 
%frozen-core full CI quality electronic energies from CAS-v2RDM calculations 
complete active space (CAS) CI quality electronic energies from CAS-v2RDM calculations
that only consider the PQG constraints. Toward this aim, we developed an ML protocol called data-driven v2RDM (DDv2RDM) that learns differences between CI and v2RDM energies using features that can be generated from RDMs optimized at the PQG level of theory. Crucial to the success of this model is the recognition that the T1 and T2 conditions are {\em partial} three-particle conditions in the sense that they are expressible in terms of the 2RDM, without knowledge of any three-body RDMs. As such, violations in these conditions can be evaluated using 2RDMs generated at the PQG level of theory. Here, we show that these intrinsic features of the v2RDM method are necessary ingredients for a transferable ML model. Moreover, this procedure paves the way for additional, improved data-driven v2RDM models that consider additional higher-order lifted constraints\cite{Mazziotti12_263002,Mazziotti23_153001} that also only depend on the 2RDM and, thus, could serve as features in such ML-based models.

The elements of the 1RDM and the 2RDM are defined in second-quantized form as
\begin{equation}
\label{EQN:D1}
    {}^1D^{p_\sigma}_{q_\sigma} = \langle \Psi | \hat{a}^\dagger_{p_\sigma}\hat{a}_{q_\sigma} |\Psi\rangle
\end{equation}
and
\begin{equation}
\label{EQN:D2}
    {}^2D^{p_\sigma q_\tau}_{r_\sigma s_\tau} = \langle \Psi | \hat{a}^\dagger_{p_\sigma}\hat{a}^\dagger_{q_\tau} \hat{a}_{ s_\tau} \hat{a}_{r_\sigma}|\Psi\rangle
\end{equation}
respectively. Here, the indices $p$, $q$, $r$, and $s$ represent spatial molecular orbital labels, and $\sigma$ and $\tau$ are spin labels. We are considering only non-relativistic Hamiltonians, in which case the non-zero blocks of ${}^1{\bf D}$ and ${}^2{\bf D}$ are spin conserving. In order for the 2RDM to be physically meaningful, it must satisfy a number of statistical conditions: it should have a fixed trace, be Hermitian,  be antisymmetric with respect to the exchange of particle labels, and contract to the 1RDM. These RDMs must also be positive semidefinite ({\em e.g.}, ${}^2{\bf D}\succeq 0$). At the PQG level of theory,  the one-hole RDM (${}^1{\bf Q}$), the two-hole RDM (${}^2{\bf Q}$), and the particle-hole RDM (${}^2{\bf G}$) should also be positive semidefinite, and the elements of these RDMs should map onto the elements of the 2RDM and 1RDM according to the anticommutation relations of fermionic creation and annihilation operators. The elements of ${}^1{\bf Q}$,  ${}^2{\bf Q}$, and ${}^2{\bf G}$ are defined as
\begin{equation}
\label{EQN:Q1}
    {}^1Q^{p_\sigma}_{q_\sigma} = \langle \Psi | \hat{a}_{p_\sigma} \hat{a}^\dagger_{ q_\sigma}|\Psi\rangle
\end{equation}
\begin{equation}
\label{EQN:Q2}
    {}^2Q^{p_\sigma q_\tau}_{r_\sigma s_\tau} = \langle \Psi | \hat{a}_{p_\sigma}\hat{a}_{q_\tau} \hat{a}^\dagger_{ s_\tau} \hat{a}^\dagger_{r_\sigma}|\Psi\rangle
\end{equation}
and
\begin{equation}
\label{EQN:G2}
    {}^2G^{p_\sigma q_\tau}_{r_\kappa s_\lambda} = \langle \Psi | \hat{a}^\dagger_{p_\sigma}\hat{a}_{q_\tau} \hat{a}^\dagger_{ s_\lambda} \hat{a}_{r_\kappa}|\Psi\rangle
\end{equation}
respectively. Here, we see that ${}^1{\bf Q}$ and ${}^2{\bf Q}$ have the same spin-block structures as ${}^1{\bf D}$ and ${}^2{\bf D}$, respectively, whereas that of ${}^2{\bf G}$ is more complex. The symbols $\kappa$ and $\lambda$ represent spin labels, and the non-zero spin blocks of ${}^2\mathbf{G}$ are those for which the number of $\alpha$-spin ($\beta$-spin) creation operators equals the number of $\alpha$-spin ($\beta$-spin) annihilation operators.

At the full 3POS level of theory, four additional three-body RDMs should be positive semidefinite and map to one another and to the 2RDM (see Ref.~\citenum{DePrince21_174110} for additional details). The elements of these RDMs are defined by 
\begin{equation}
\label{EQN:D3}
    ^3D^{p_\sigma q_\tau r_\kappa}_{s_\sigma t_\tau u_\kappa}=
              \langle \Psi |
              \hat{a}^\dagger_{p_\sigma}  \hat{a}^\dagger_{q_\tau} \hat{a}_{r_\kappa}^{\dagger}
              \hat{a}_{u_\kappa}  \hat{a}_{t_\tau}  \hat{a}_{s_\sigma} 
              | \Psi \rangle,
\end{equation}
\begin{equation}                                                                                             \label{EQN:E3}
    ^3E^{p_\sigma q_\tau r_\kappa}_{ s_\lambda t_\mu u_\nu} =
              \langle \Psi |
              \hat{a}^\dagger_{p_\sigma}  \hat{a}^\dagger_{q_\tau} \hat{a}_{r_\kappa}
              \hat{a}_{u_\nu}^{\dagger}  \hat{a}_{t_\mu}  \hat{a}_{s_\lambda} 
              | \Psi \rangle,
\end{equation}
\begin{equation}                                                                                             \label{EQN:F3}
    ^3F^{p_\sigma q_\tau r_\kappa}_{ s_\lambda t_\mu u_\nu} =
            \langle \Psi |
            \hat{a}_{u_\nu}^{\dagger}  \hat{a}_{t_\mu}  \hat{a}_{s_\lambda}
            \hat{a}^\dagger_{p_\sigma}  \hat{a}^\dagger_{q_\tau} \hat{a}_{r_\kappa}
            | \Psi \rangle,
\end{equation}
and
\begin{equation}                                                                                             \label{EQN:Q3}
      ^3Q^{p_\sigma q_\tau r_\kappa}_{ s_\sigma t_\tau u_\kappa}=
              \langle \Psi |
              \hat{a}_{p_\sigma}  \hat{a}_{q_\tau} \hat{a}_{r_\kappa}
              \hat{a}^\dagger_{u_\kappa}  \hat{a}^\dagger_{t_\tau}  \hat{a}^{\dagger}_{s_\sigma} 
              | \Psi \rangle.
\end{equation}
Here, $\mu$, and $\nu$ represent spin functions, and, as was the case for ${}^2{\bf G}$, the non-zero spin blocks of ${}^3\mathbf{E}$ and ${}^3\mathbf{F}$ are those for which the number of $\alpha$-spin ($\beta$-spin) creation operators equals the number of $\alpha$-spin ($\beta$-spin) annihilation operators.  One can also define weaker three-body constraints on the 2RDM by taking specific linear combinations of the three-particle RDMs defined above.\cite{Erdahl78_697, Percus04_2095} In particular, we have
\begin{align}
    \label{EQN:T1}
    \mathbf{T1} = {}^3\mathbf{D} + {}^3\mathbf{Q}\\
    \label{EQN:T2}
    \mathbf{T2} = {}^3\mathbf{E} + {}^3\mathbf{F}.
\end{align}
Both of these RDMs should be positive semidefinite, and, importantly, the right-hand sides of Eqs.~\ref{EQN:T1} and \ref{EQN:T2} can be defined without knowledge of any three-body RDM. As a result, we can evaluate errors in the T1 and T2 conditions (the appearance of negative eigenvalues in these matrices) using RDMs optimized at the PQG level of theory.

Our aim is to develop an ML model to learn the difference between CI energies and those generated at the PQG level of theory using features based on the RDMs defined above. In order to generate a viable and transferable model, we performed extensive exploratory data analysis and feature engineering. We include the following features in the DDv2RDM model: 
the entropy of one- and two-body RDMs,
\begin{equation}                                                           \label{EQN:entropy}
  S=\sum_{i} n_{i} \log_{2} n_{i}
\end{equation}
where $n_{i}$ are the eigenvalues of the RDM,
traces of the spin blocks of the cumulant 2RDM, Tr$({}^2\Delta)$, with 
\begin{equation}                                                           \label{EQN:cumulant}
  {}^2\Delta = {}^2\mathbf{D} - {}^1\mathbf{D}\wedge {}^1\mathbf{D},
\end{equation}
the norms of the spin blocks of the cumulant 2RDM, $|| ^{2}{\Delta}||^{2}$, and
information related to violations in the T1 and T2 conditions ({\em i.e.}, the existence of negative eigenvalues). We quantify both the frequency of the T1 and T2 violations (how many eigenvalues are negative, as a percentage), as well as the magnitude and distribution of these violations (average, variance, and root mean square of the negative eigenvalues).

Our data set includes these features computed along potential energy curves for 36 neutral diatomic species formed from hydrogen, lithium, beryllium, boron, carbon, nitrogen, oxygen, and fluorine atoms. We consider multiple spin states, {\em i.e.}, singlet and triplet states for molecules with even numbers of active electrons, and doublet and quartet states for molecules with odd numbers of active electrons. Dissociation curves are evaluated over inter-atomic distances ranging from 0.6 \AA--2.9 \AA~for all cases.
%; expanded ranges including  3.0 \AA--3.5 \AA~were used in some cases. 
We also considered multiple basis sets (STO-3G, 6-31G, and cc-pVDZ). For minimal basis (STO-3G) calculations, the active space was chosen to be the full orbital space, whereas correlated calculations performed within the 6-31G and cc-pVDZ basis sets considered only the full valence space to be active. 
%The case of molecular hydrogen is an exception, where we  use the full space in all basis sets. 
Canonical restricted (open-shell) Hartree-Fock orbitals were used in all correlated calculations ({\em i.e.}, the orbitals were not further optimized for v2RDM- or CI-driven CASSCF). All v2RDM calculations account for PQG\cite{Percus64_1756} ensemble $N$-representability constraints. The v2RDM calculations were considered converged when the primal-dual energy gap fell below 10$^{-6}$ $E_{\rm h}$ and the primal/dual errors fell below 10$^{-5}$. We refer the reader to Ref.~\citenum{DePrince19_6164} for a description of these quantities.
We used a value of $-1 \times 10^{-8}$ for eigenvalues of T1 and T2 as a cutoff when evaluating the statistical measures of violations in these conditions. 

\input{results}

%\section{Conclusions}
%\input{conclusions}

To summarize, we have introduced a data-driven v2RDM-based method, DDv2RDM, which is an ML model for the efficient recovery of CI-quality electronic energies.
Using a diverse dataset composed of diatomic species in multiple spin states and basis sets,  our models exhibit mean RMSEs of 1.4443$\times 10^{-3}$ to 1.7841$\times 10^{-3}$  $E_{\rm h}$, which is near chemical accuracy ({\em i.e.}, 1.5936$\times 10^{-3}$ $E_{\rm h}$ or 1 kcal/mol) and a significant improvement over the native accuracy of v2RDM calculations performed under two-body $N$-representability conditions. A crucial component of the success of this model is the use of features based on high-order $N$-representability conditions (T2) that can be evaluated with knowledge RDMs that can be optimized efficiently using low-order v2RDM theory.
We also introduced SHAP value analysis, a feature importance method based on cooperative game theory, which can provide insight into the physical information included in the feature set and the impact of these features on the ability of the model to learn the v2RDM energy correction.
The insight provided by this method confirmed our expectation that violations in the T2 condition carry important information that could be exploited by the ML model, while also revealing the surprising impact of the entropy of the one-hole RDM on the model. This work paves the way for improved ML models that refine energy estimates from v2RDM theory and that can provide high-accuracy alternatives to intractable CI-based calculations on large numbers of strongly correlated electrons.

%In followup studies, we will expand this method by exploring larger systems, beyond diatomics and incorporate the effects of the feature importance analysis into our models.

%\section{Computational Details}

\vspace{0.5cm}

{\bf Supporting Information} Energies for different spin states, model parameters, error analysis, dissociation curves and corrections for each diatomic, and SHAP details. 

\vspace{0.5cm}

\begin{acknowledgments}This material is based upon work supported by the U.S. Department of Energy, Office of Science, Office of Advanced Scientific Computing Research and Office of Basic Energy Sciences, Scientific Discovery through the Advanced Computing (SciDAC) program under Award No. DE-SC0022263.
This material is based on work supported by the National Science Foundation (NSF CAREER Award) under Grant CHE-2143354 (G. M. J. and K. D. V.). \\ 
\end{acknowledgments}

%\noindent {\bf DATA AVAILABILITY}\\

%    The data that support the findings of this study are available from the corresponding author upon reasonable request.

\bibliography{bib/Journal_Short_Name,bib/main,bib/deprince,bib/rdm}

\end{document}

%% file: results.tex
\label{SEC:RESULTS}

\begin{figure}[!htpb]
    \centering
    \begin{subfigure}{\linewidth}
        \centering
        \includegraphics[width=\linewidth]{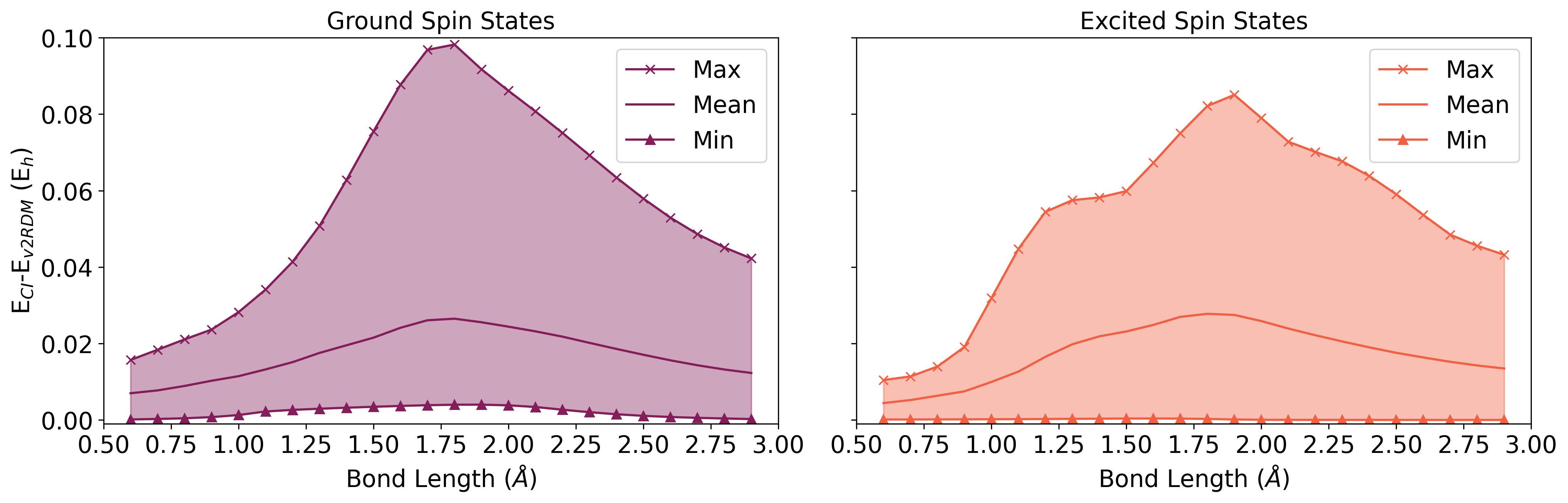}
        \caption{}
        \label{fig:STO3Gallspread}
    \end{subfigure}
    \hfill
    \begin{subfigure}{\linewidth}
        \centering
        \includegraphics[width=\linewidth]{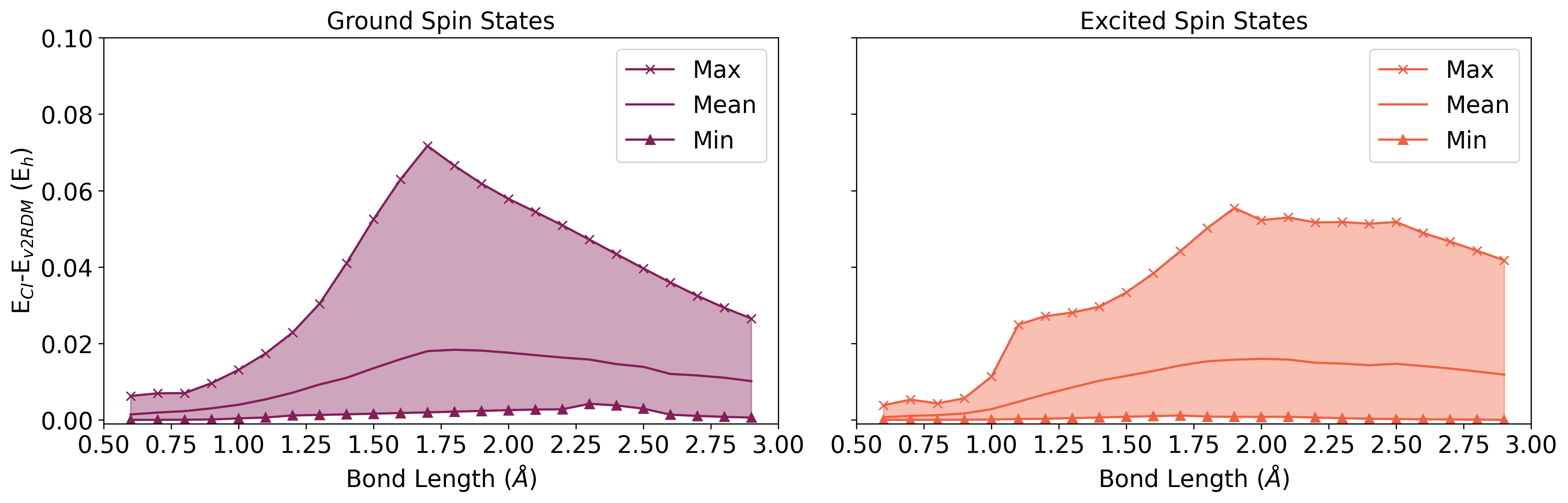}
        \caption{}
        \label{fig:631Gallspread}
    \end{subfigure}
    \hfill
    \begin{subfigure}{\linewidth}
        \centering
        \includegraphics[width=\linewidth]{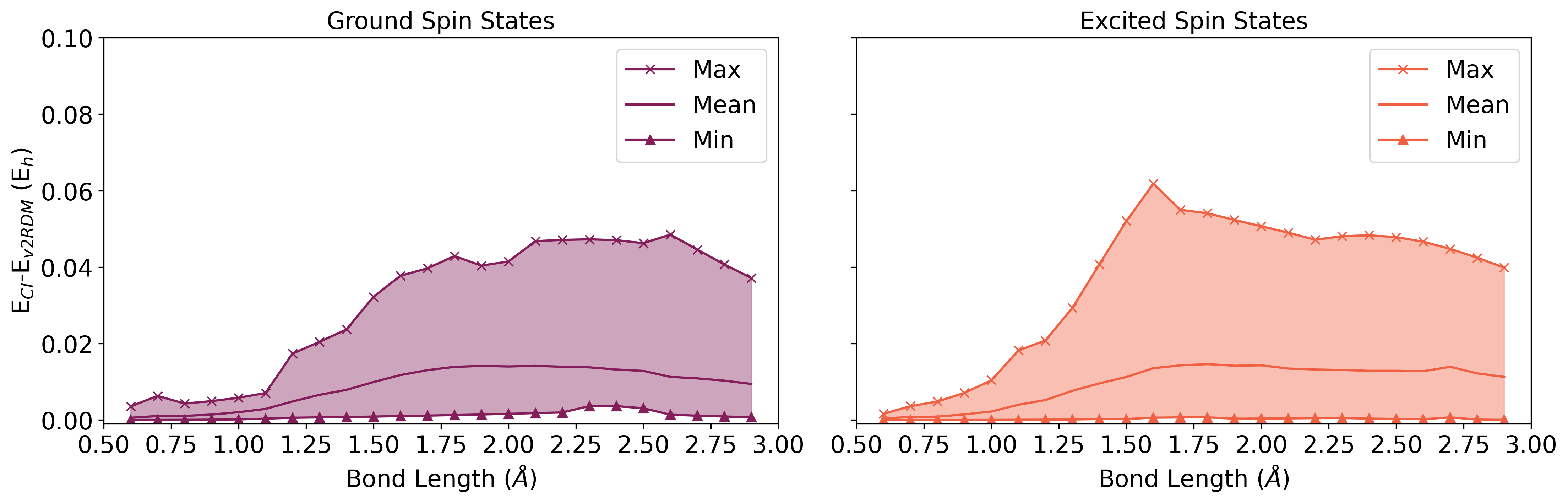}
        \caption{}
        \label{fig:ccpVDZallspread}
    \end{subfigure}
    \hfill    
    
    \caption{The minimum, mean, and maximum values of the target values ($E_{\text{CI}}-E_{\text{v2RDM}}$) at each bond length for ground (left) and excited (right) spin states optimized within the (a) STO-3G, (b) 6-31G, and (c) cc-pVDZ basis sets.}
    \label{fig:energy_spreads}
\end{figure}

In our ML model, we formulate the target value as a correction to the v2RDM energy, {\em i.e.}, the difference between CI and v2RDM energies, $E_{\text{CI}}-E_{\text{v2RDM}}$.
Systems for which the mean of the root-mean square errors (RMSEs) of the T1 or T2 violations across the potential energy curve are less than 10$^{-6}$ are excluded.
Since we consider multiple spin states, the ML model is evaluated on two sets of data, corresponding to systems in their ground or excited spin states, as determined by the CI energy at the equilibrium geometry.
The spin multiplicities for all systems can be found in the Supporting Information (Fig.~\ref{fig:spin_multiplicities}).

To gain insights into the distributions of the target values, we analyze the minimum, mean, and maximum values across the potential energy curves for all molecules represented within the STO-3G, 6-31G, and cc-pVDZ basis sets (Fig.~\ref{fig:STO3Gallspread}, \ref{fig:631Gallspread}, and \ref{fig:ccpVDZallspread}, respectively).
As shown in Figs.~\ref{fig:STO3Gallspread}, \ref{fig:631Gallspread}, and \ref{fig:ccpVDZallspread}
the maximum deviation often occurs at intermediate bond lengths, with the exception of the ground spin states optimized in the cc-pVDZ basis [left-hand panel of Fig.~\ref{fig:ccpVDZallspread}]. In this case, the maximum deviation occurs closer to the bond dissociation limit.
For the ground spin states, the maximum deviation between v2RDM and CI energies occurs for the singlet state of the C$_2$ molecule, regardless of the basis set. For the STO-3G, 6-31G, and cc-pVDZ basis sets, these maximum deviations are 0.0982 $E_{\text{h}}$, 0.0717 $E_{\text{h}}$, and 0.0485 $E_{\text{h}}$, respectively. 
%For the STO-3G basis set, the maximum deviation is 0.0982 $E_{\text{h}}$ for the singlet state of C$_{2}$ at 1.8 \AA~(Fig.~\ref{fig:STO3Gallspread}, left), for 6-31G is 0.0717 $E_{\text{h}}$ for the singlet state at 1.7 \AA~(Fig.~\ref{fig:631Gallspread}, left), and for cc-pVDZ is 0.0485 $E_{\text{h}}$ for the triplet state at 2.6 \AA~(Fig.~\ref{fig:ccpVDZallspread}, left).
For the excited spin states, the maximum deviation between v2RDM and CI energies occurs for the doublet state of the BC molecule in the STO-3G and 6-31G basis sets and the singlet state of the BN molecule in the cc-pVDZ basis set. For the STO-3G, 6-31G, and cc-pVDZ basis sets, these maximum deviations are 0.0850 $E_{\text{h}}$, 0.0554 $E_{\text{h}}$, and 0.06179 $E_{\text{h}}$, respectively. 
%The maximum deviations of the excited spin states in each basis are 0.0850 $E_{\text{h}}$ for the doublet state of the BC molecule at 1.9 \AA~(STO-3G, Fig.~\ref{fig:STO3Gallspread}, right), 0.0554 $E_{\text{h}}$ for the doublet state of  BC at 1.9 \AA~(6-31G, Fig.~\ref{fig:631Gallspread}, right), and 0.06179 $E_{\text{h}}$ for the singlet state of BN at 1.6 \AA~(cc-pVDZ, Fig.~\ref{fig:ccpVDZallspread}, right).
Overall, these large deviations between CI and v2RDM energies motivate the development of an ML-based methodology to correct the v2RDM energies.

All ML models developed for this study utilize kernel ridge regression (KRR) with a radial basis function (RBF) kernel, defined as 
\begin{equation}
    k(x_{i},x_{j}) = \exp{\left( -\gamma || x_{i}-x_{j} ||^{2}_{2} \right)},
    \label{eq:rbf_kernel}
\end{equation}
where $\gamma$ is a hyperparameter.
In all models, input features were normalized using a MinMaxScaler function, while $\gamma$ and the model regularization parameter $\alpha$ were optimized using a five-fold parameter grid search as implemented in Sci-kit Learn.\cite{Pedregosa2011} 
An example of the model parameters examined can be found in the Supporting Information (Section~\ref{section:model_parameters}).

\begin{figure}[H]
     \centering
     \begin{subfigure}[b]{1\linewidth}
         \centering
         \includegraphics[width=0.75\linewidth]{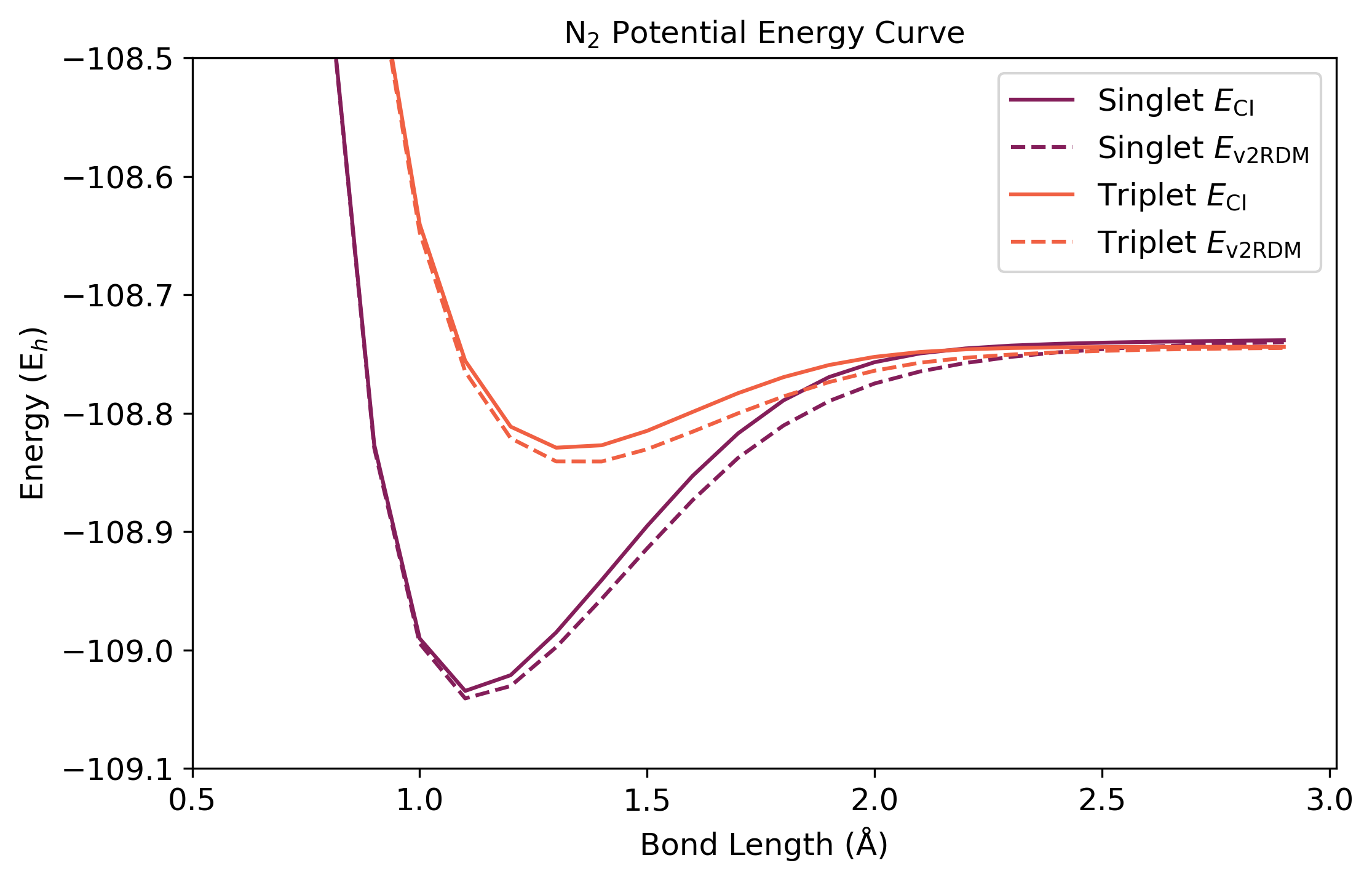}
         \caption{}
         \label{fig:cc-pVDZ_PEC_NN}
     \end{subfigure}
     \hfill
     \begin{subfigure}[b]{1\linewidth}
         \centering
         \includegraphics[width=\linewidth]{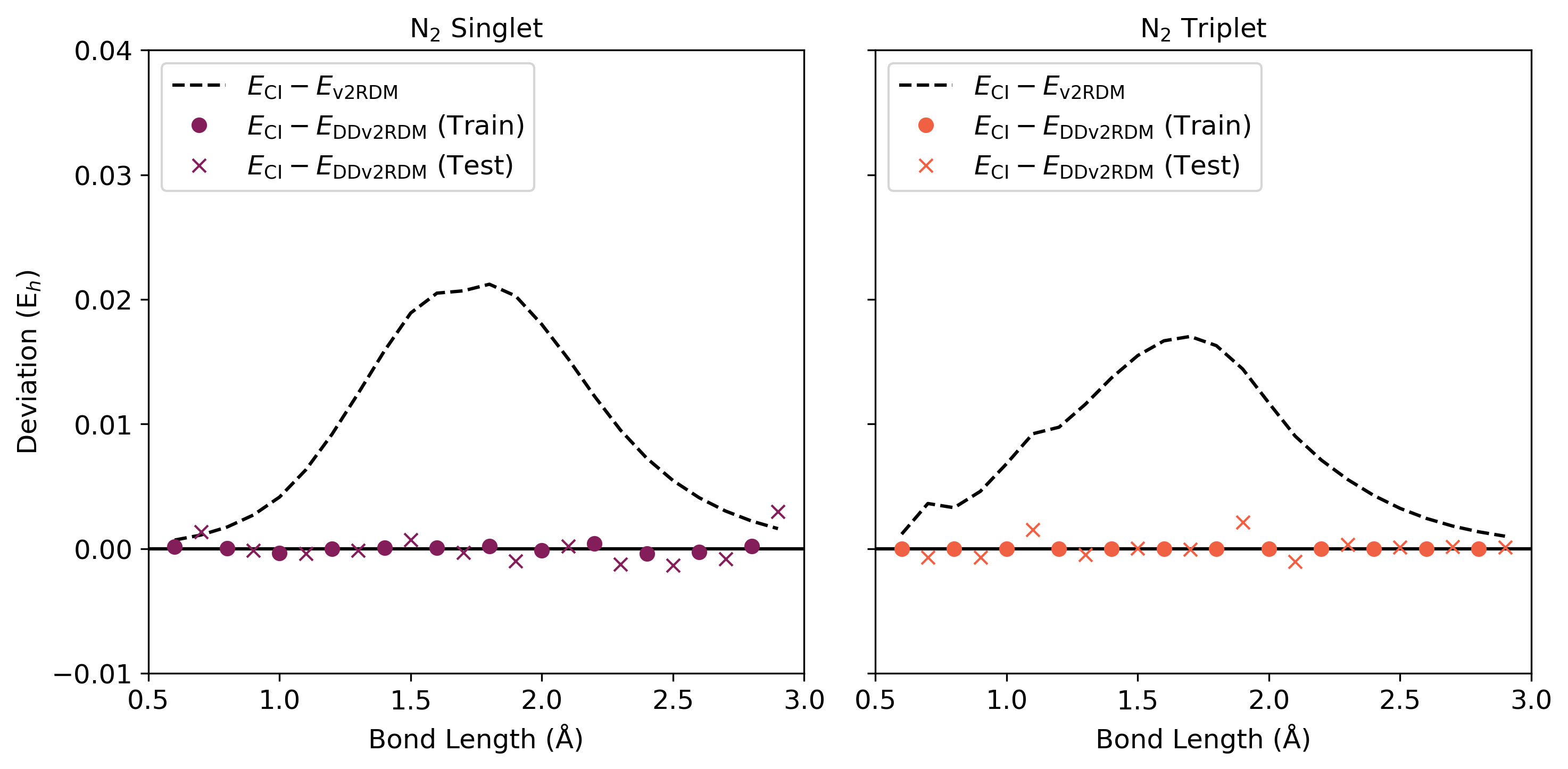}
         \caption{}
         \label{fig:ccpvdzstackedNN}
     \end{subfigure}
        \caption{(a) The potential energy curves of N$_{2}$ using the cc-pVDZ basis set for the ground (singlet, purple) and excited  (triplet, orange) spin states. The CI energy is shown as a solid line and the v2RDM energy as a dashed line.
        (b) Deviations between CI and (DD)v2RDM energies. %The difference between $E_{\text{CI}}$ and $E_{\text{v2RDM}}$ are shown using a dashed black line, $E_{\text{CI}}$ using a solid black line, and the difference between $E_{\text{CI}}$ and the DDv2RDM correction ($\delta E_{\text{DDv2RDM}}$) are shown in purple and orange for the ground and excited states, respectively. 
        %The training set for both DDv2RDM corrections are denoted using a circle and the test set as a cross.
        }
        \label{fig:ccpvdzNN}
\end{figure}

Figure \ref{fig:ccpvdzNN} shows CI and v2RDM potential energy curves [panel a], as well as deviations between CI energies and those from v2RDM and DDv2RDM [panel b] for a representative example: the N$_2$ molecule within the cc-pVDZ basis set.
%We have selected the ground and excited states of N$_{2}$ together with the cc-pVDZ basis set as a proof-of-concept of the proposed methodology.
%The results from the two individual models trained separately for each spin state are shown in Figure \ref{fig:cc-pVDZ_PEC_NN}.
Here, we consider evenly spaced points along the curves, with 50\% of the data used for training and the remaining data reserved for testing.
%, which results in less over-fit models than 
%Evenly spreadout points along the potential energy curve, with 50\% of the available data used for training and the remaining data for testing provided less overfit models than traditional train/test splits, such as 
%an 80\%/20\% split (Supporting Information Section \ref{section:individual_curves}). 
Figure \ref{fig:ccpvdzNN}b shows that the maximum deviation between CI and v2RDM for the singlet state is 2.1222$\times 10^{-2}$ $E_{\text{h}}$ at 1.8 \AA,~while that for the triplet state is 1.7031$\times 10^{-2}$ $E_{\text{h}}$ at 1.7 \AA.
Overall, the RMSE between CI and v2RDM energies are 1.2149$\times 10^{-2}$ $E_{\text{h}}$ and 9.6136$\times 10^{-3}$ $E_{\text{h}}$ for the ground and excited states, respectively.
Addition of the DDv2RDM correction reduces the RMSE of the singlet spin state to 2.3550$\times 10^{-4}$ $E_{\text{h}}$ and 1.1683$\times 10^{-3}$ $E_{\text{h}}$ for the training and test sets, respectively. 
For the triplet spin state, the training set RMSE is reduced to 7.0262 $\times 10^{-9}$ $E_{\text{h}}$, while for the test set the error is reduced to 8.8469 $\times 10^{-4}$ $E_{\text{h}}$.
These reductions translate into percent improvements of 98.1\% and 90.3\% for the training and testing for the singlet state, respectively, while for the triplet state, these improvements are 99.9\% and 90.8\%, respectively.
%By using the percent improvements over the initial deviation, this example highlights how the ML derived correction can improve the accuracy of the v2RDM method using ML.

\begin{figure}[!htpb]
     \centering
     \begin{subfigure}[b]{1\linewidth}
         \centering
         \includegraphics[width=\linewidth]{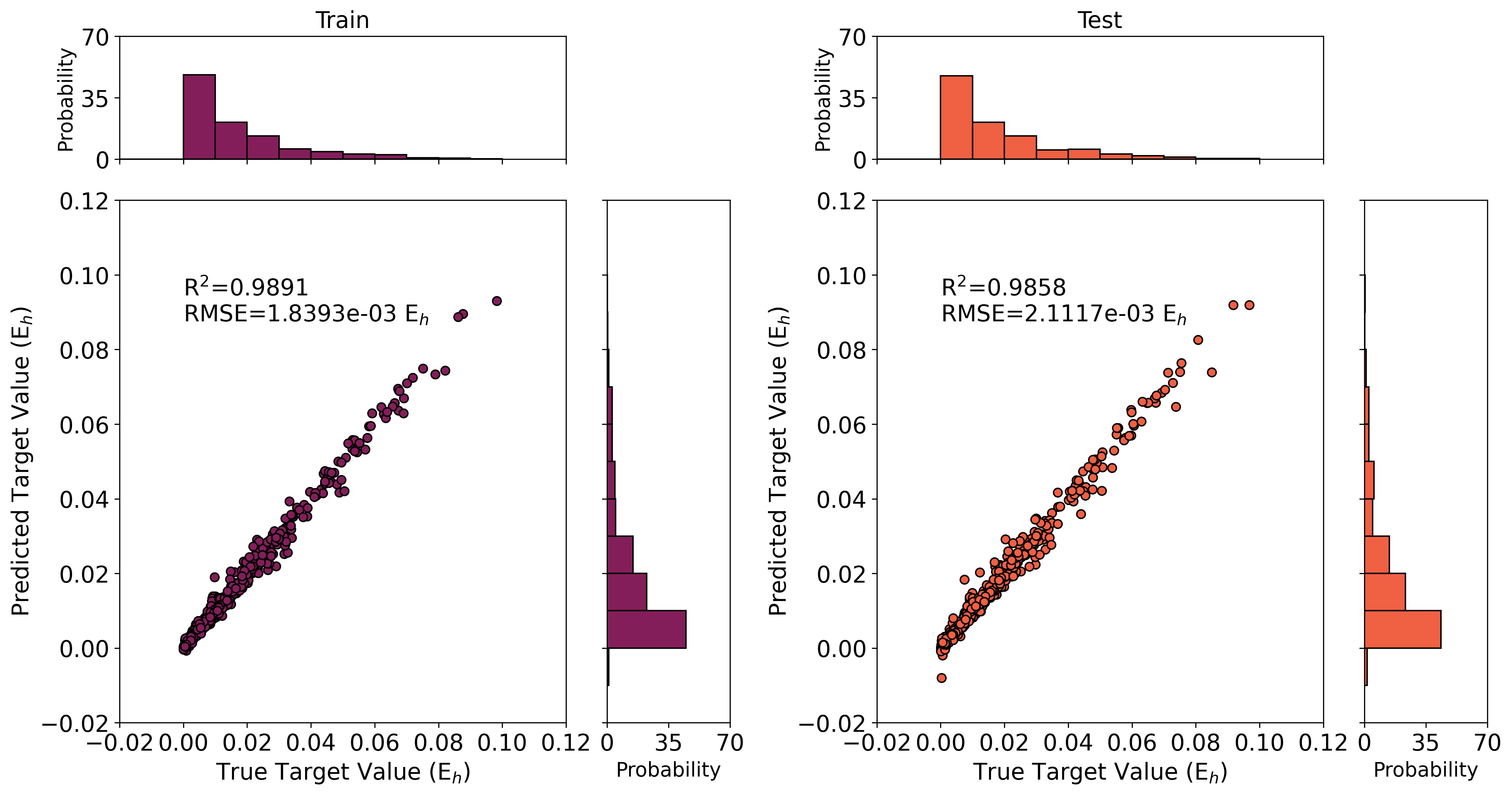}
         \caption{}
         \label{fig:STO-3G_bigstack_error}
     \end{subfigure}
     \hfill
     \begin{subfigure}[b]{1\linewidth}
         \centering
         \includegraphics[width=\linewidth]{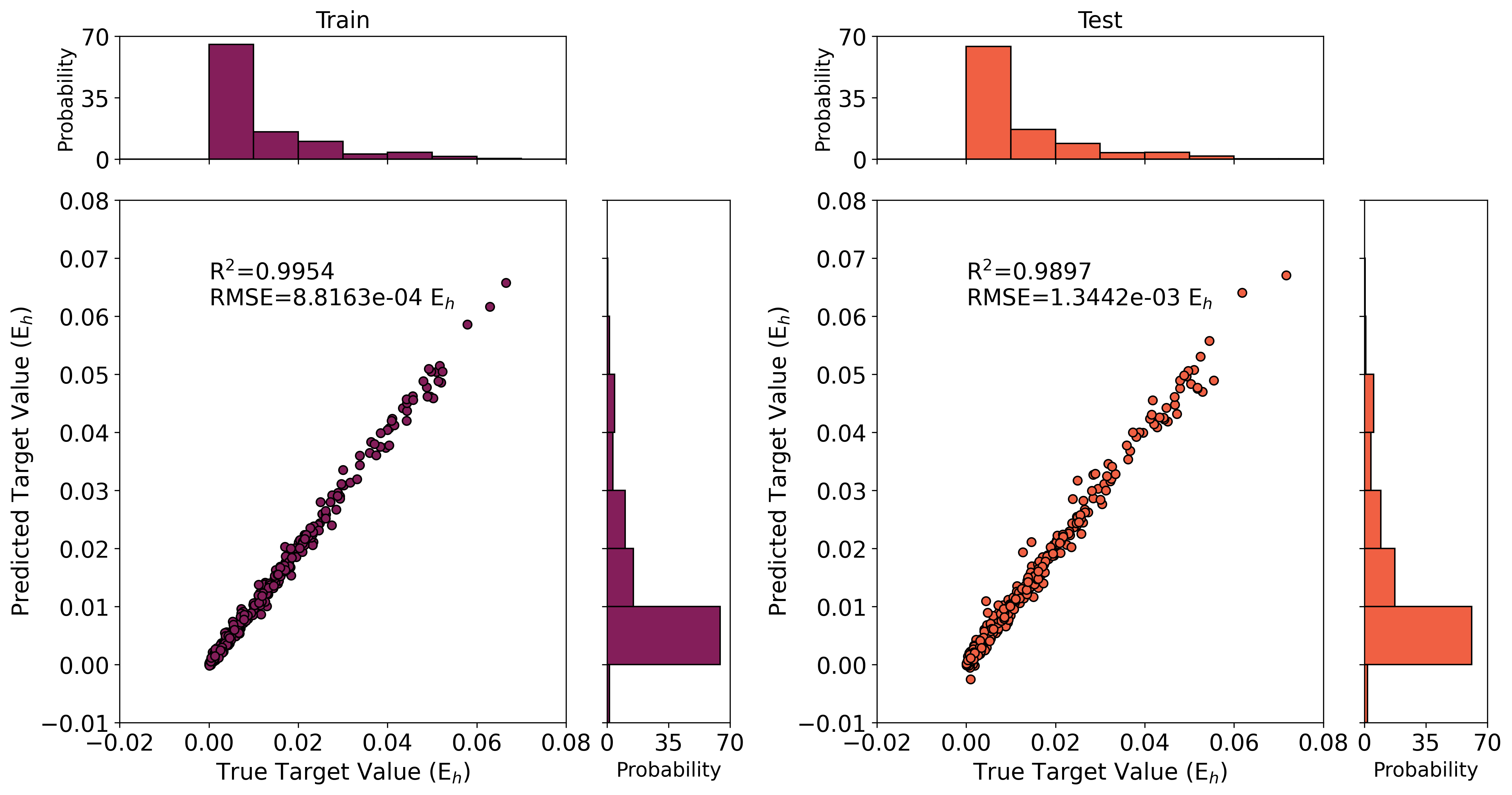}
         \caption{}
         \label{fig:6-31G_bigstack_error}
     \end{subfigure}
     \hfill
     \begin{subfigure}[b]{1\linewidth}
         \centering
         \includegraphics[width=\linewidth]{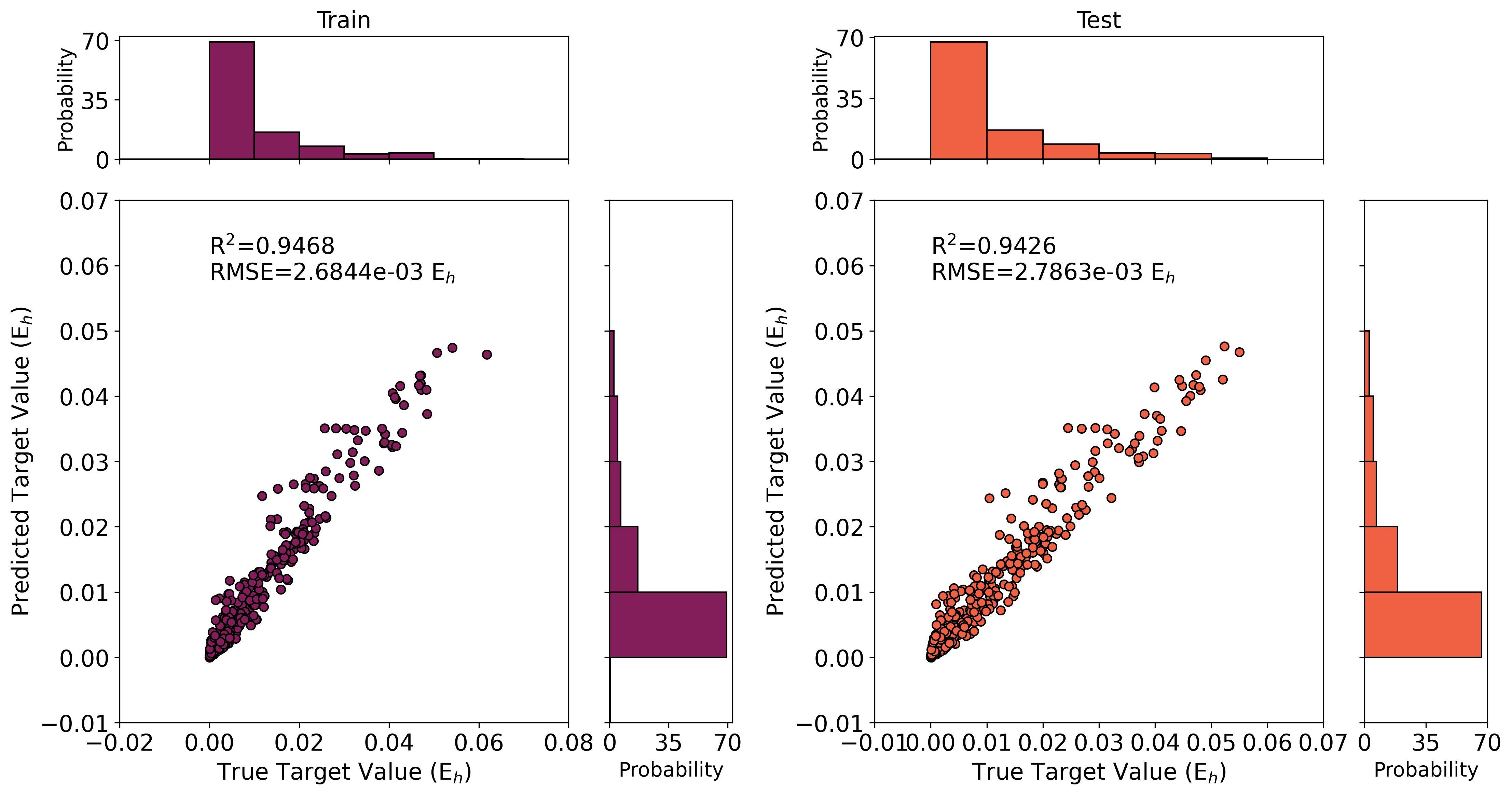}
         \caption{}
         \label{fig:cc-pVDZ_bigstack_error}
     \end{subfigure}     
        \caption{Regression parity plots comparing the true, calculated target values with the predicted target values for (a) STO-3G, (b) 6-31G, and (c) cc-pVDZ basis sets. Probability density histograms are given on the x- and y-axis. The left (purple) panel shows the training and right (orange) panel shows the test data.}
        \label{fig:bigstack_error}
\end{figure}

We have assessed the transferability of the DDv2RDM method for two cases: (i) different models were developed for ground versus excited spin states within a given basis set and (ii) a unified model was created that considered all available data within a given basis set. % utilizing data for different diatomic molecules on individual spin states (ground or excited spin state) or by using both sets of data.
Based on the T1/T2 selection criteria, the dataset for the ground spin states includes 22 STO-3G, 17 6-31G, and 18 cc-pVDZ systems, while that for the excited spin states includes 19 STO-3G, 21 6-31G, and 19 cc-pVDZ systems.
We evaluate the performance of the models using standard regression metrics, which include the RMSE and the coefficient of determination (R$^{2}$).
It was found that the most transferable and least overfit model was obtained when both sets of data were used for training and testing DDv2RDM (Section \ref{section:joint_curves} in the Supporting Information).
The regression parity plots for this model are shown in Fig.~\ref{fig:bigstack_error} for the three basis sets under consideration.
For the STO-3G basis set (Fig.~\ref{fig:STO-3G_bigstack_error}), the model achieves R$^{2}$ values of 0.9891 and 0.9858 for the train and test set, respectively.
The training set has an RMSE of 1.8393$\times 10^{-3}$ $E_{\text{h}}$ and the test set has an RMSE of 2.1117$\times 10^{-3}$ $E_{\text{h}}$.
Figure \ref{fig:6-31G_bigstack_error} shows that the model trained on 6-31G data exhibits increased accuracy with respect to the STO-3G basis set.
Training and testing R$^{2}$ values are 0.9954 and 0.9897, respectively,  which correspond to RMSE values of 8.8163$\times 10^{-4}$ $E_{\text{h}}$ and 1.3442$\times 10^{-3}$ $E_{\text{h}}$, respectively.
The least overfit model is the cc-pVDZ model (Figure \ref{fig:cc-pVDZ_bigstack_error}), which has R$^{2}$ values of 0.9468 and 0.9426 for training and testing, respectively, with RMSEs of 2.6844$\times 10^{-3}$ $E_{\text{h}}$ and 2.7863$\times 10^{-3}$ $E_{\text{h}}$, respectively.
%When compared to the other models that are extensively discussed in the Supporting Information Sections \ref{section:individual_curves} and \ref{section:joint_curves}, there is a slight increase in error for the training set but overall, this model exhibits less overfitting than the models trained on individual curves or trained on all curves of a certain energy state.

\begin{figure}[!htpb]
     \centering
     \begin{subfigure}[b]{1\linewidth}
         \centering
         \includegraphics[width=\linewidth]{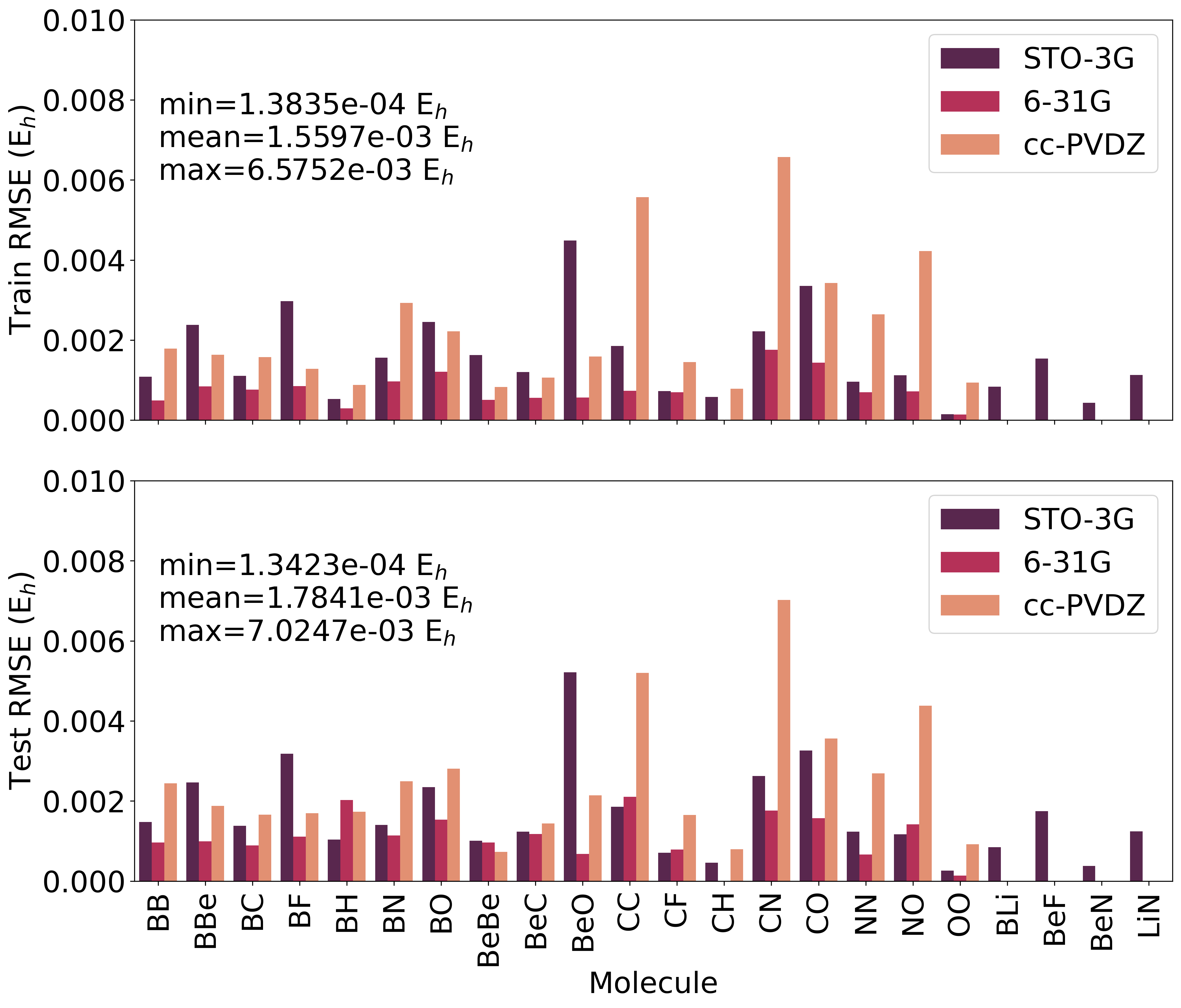}
         \caption{}
         \label{fig:both_GS_regression_stats}
     \end{subfigure}
     \hfill
     \begin{subfigure}[b]{1\linewidth}
         \centering
         \includegraphics[width=\linewidth]{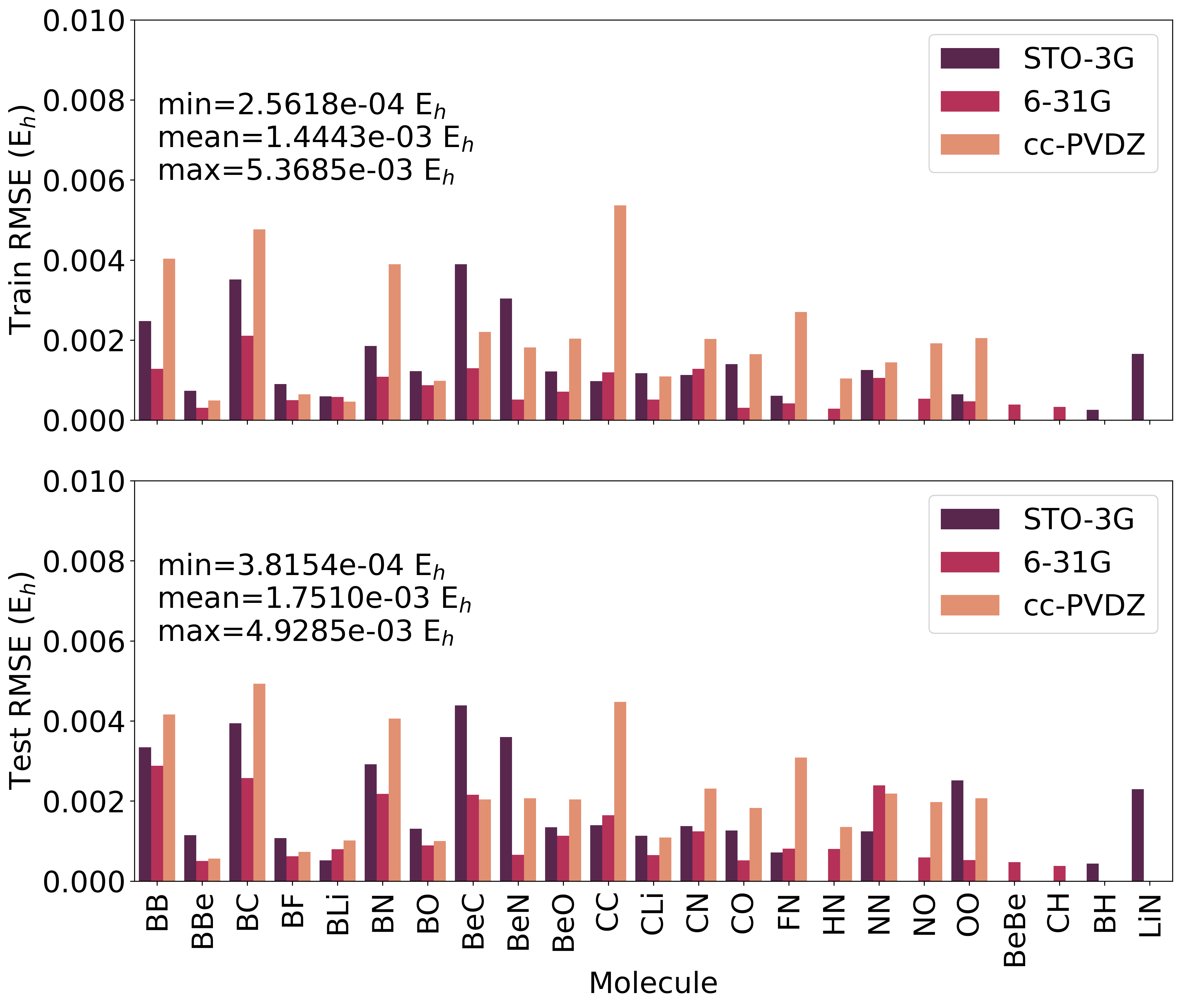}
         \caption{}
         \label{fig:both_ES_regression_stats}
     \end{subfigure}
        \caption{Error plots for the model trained considering all available data within a given basis set where the (a) ground and (b) excited spin states are partitioned into the training (top; 50\% of the data) and test set (bottom; 50\% of the data).}
        \label{fig:both_all_regression_stats}
\end{figure}

We further examine the performance of our model by analyzing the RMSE of the DDv2RDM energy for each system in the ground and excited spin states, as shown in Fig.~\ref{fig:both_all_regression_stats}.
For the STO-3G basis set, the molecules with the largest training and test errors are the singlet (ground) state of BeO (4.4841$\times 10^{-3}$ $E_{\text{h}}$ and 5.2121$\times 10^{-3}$ $E_{\text{h}}$, respectively, Fig.~\ref{fig:both_GS_regression_stats}) and the singlet (excited) state of BeC (3.8980$\times 10^{-3}$ $E_{\text{h}}$ and 4.3875$\times 10^{-3}$ $E_{\text{h}}$, respectively, Fig.~\ref{fig:both_ES_regression_stats}).
The largest RMSEs for the 6-31G basis set correspond to four different systems for the training and test sets for both spin states.
For the ground state, shown in Fig.~\ref{fig:both_GS_regression_stats}, the largest error in the training set corresponds to the doublet state of CN with an RMSE of 1.7559$\times 10^{-3}$ $E_{\text{h}}$ and, for the test set, the singlet state of C$_{2}$ with an RMSE of 2.1021$\times 10^{-3}$ $E_{\text{h}}$.
The excited spin states have a training RMSE of 2.1099$\times 10^{-3}$ $E_{\text{h}}$ for the doublet state of BC and test RMSE of 2.8779$\times 10^{-3}$ $E_{\text{h}}$ for the singlet state of B$_{2}$.
Among the ground state systems, CN has the largest training and test RMSE of 6.5752$\times 10^{-3}$ $E_{\text{h}}$ and 7.0247$\times 10^{-3}$ $E_{\text{h}}$, respectively, both of which correspond to the doublet state of CN at the cc-pVDZ basis set.
Similar to the ground state systems, the largest training and test errors for the excited states correspond to two cc-pVDZ systems, where the training RMSE is 5.3685$\times 10^{-3}$ $E_{\text{h}}$ for the singlet state of C$_{2}$ and test RMSE of 4.9285$\times 10^{-3}$ $E_{\text{h}}$ for the doublet state of BC.

Lastly, we explore the issue of interpretability by investigating how the feature set affects the DDv2RDM model performance.
For that purpose, we have used SHapley Additive exPlanation (SHAP) values.\cite{NIPS2017_7062}
The SHAP analysis provides feature explanation values based on cooperative game theory that show how features (players in the game) affect the ML model in an additive manner with regards to the prior expectation value $E[f(X)]$, where $f(X)$ is the model target value dependent on the feature set $X$.
As a method based on game theory, the base value of $E[f(X)]$ corresponds to the model trained without any features (no players being present), and then features are added into the model such that the SHAP values of individual features always sum to the difference between all features (players) and no features (players) being present.
In other words, the SHAP values will sum up to the difference between $E[f(X)]$ and $f(X)$.
Based on this formulation, SHAP values can provide a clear measure of how strongly a feature positively or negatively impacts the model as each SHAP value is summed from $E[f(X)]$ to the predicted value $f(X)$.

\begin{figure}[!htpb]
     \centering
     \begin{subfigure}[b]{1\linewidth}
         \centering
         \includegraphics[width=\linewidth]{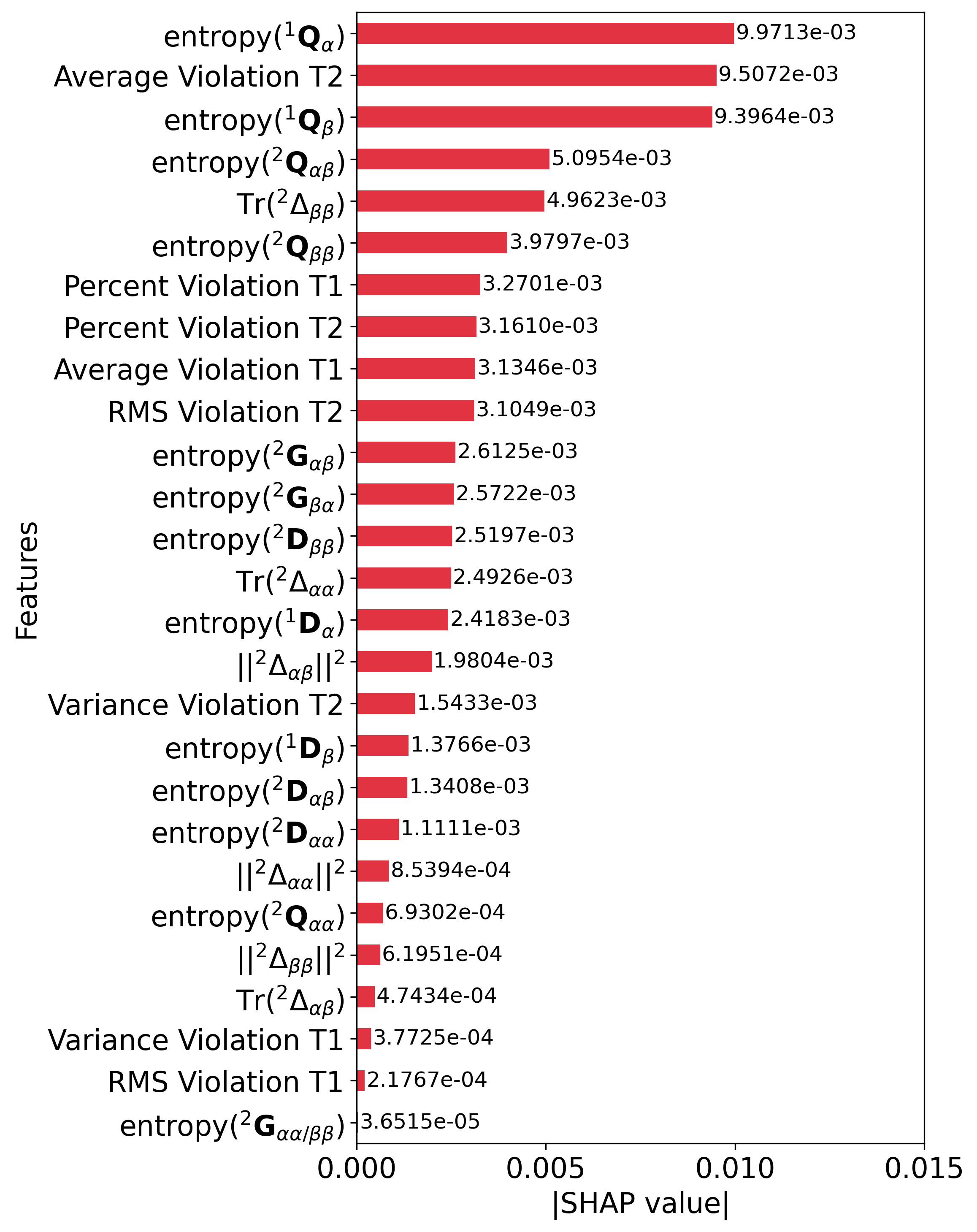}
         \caption{}
         \label{fig:6-31Gbarall_C2_1.7}
     \end{subfigure}
     \hfill
     \begin{subfigure}[b]{1\linewidth}
         \centering
         \includegraphics[width=\linewidth]{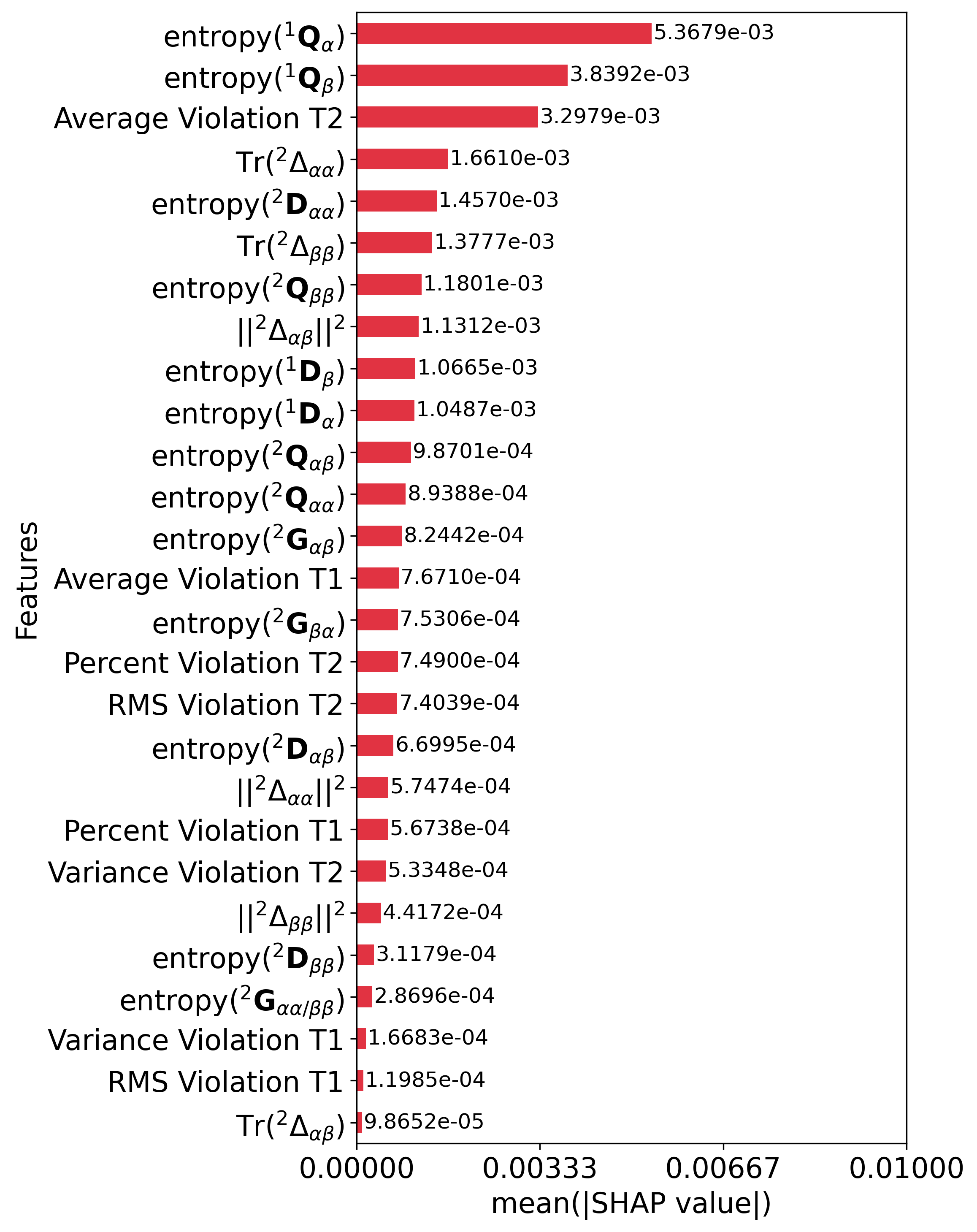}
         \caption{}
         \label{fig:6-31Gbarall}
     \end{subfigure}     
    \caption{SHAP values in the form of bar graphs for (a) the molecule with the maximum deviation (ground state of C$_{2}$ with the 6-31G basis set and at 1.7 \AA, and (b) for all the molecules obtained with the 6-31G basis set.}
    \label{fig:CC_singlet_doublet_GS_STO3G}
\end{figure}

As a measure of feature importance, we use the absolute value of the SHAP values when analyzing single systems and the mean absolute SHAP values when analyzing sets of data.
As an example, using the 6-31G basis set, we compare the absolute SHAP values of the ground spin state of the C$_{2}$ molecule at 1.7 \AA~(Fig. \ref{fig:6-31Gbarall_C2_1.7}) and the mean absolute SHAP values of the full dataset~(Fig. \ref{fig:6-31Gbarall}) consisting of 17 molecules in their ground spin state and 21 molecules in their excited spin state.
For both cases, the three most important features correspond to the average violation of the T2 conditions and the entropies of the $\alpha$- and $\beta$-blocks of the one-hole RDM ($^{1}\mathbf{Q}$).
The essential role of T2 is not surprising. Violations in the T1 and T2 conditions are the only features that are directly related to energy errors (in the sense that, for an exact solution, the T1 and T2 conditions would not be violated), and T2 is known to be the stronger condition. On the other hand, the importance of the entropy of the one-hole RDM was not anticipated, nor was the fact that the entropy of the 1RDM apparently contains much less valuable information for the model. $^{1}\mathbf{D}$ and $^{1}\mathbf{Q}$ have a complementary structure, so it is not immediately obvious why one of these RDMs would contain more useful information than the other. Nevertheless, the SHAP values clearly indicate that the entropy of $^{1}\mathbf{Q}$ significantly impacts model performance. 

%The common features that are the least important in both models are the traces of the $\alpha\beta$ blocks of the cumulant of the 2RDM [Tr$(^{2}\Delta_{\alpha\beta})$], the variance and the RMSE of the T1 violations.
Among the features that have less of an impact on model performance, a few are worth highlighting. First, as Fig.~\ref{fig:6-31Gbarall_C2_1.7} and \ref{fig:6-31Gbarall} show, the variance of the violations of the T1 condition and the RMS violations in the T1 condition are clearly not important. The average violation in the T1 condition is carries slightly more weight in the model, but this feature is still far less impactful than the entropy of ${}^1{\bf Q}$ or the average violation in the T2 condition. While violations in the T1 condition imply that the 2RDM is not $N$-representable and, thus, that the associated energy is a lower-bound to the CI energy, it is well known that T1 is a weak condition, compared to T2. From this point of view, it is not too surprising that violations in T1 do not carry much important information for the model. 
Second, for the full set of data in the 6-31G basis set [Fig.~\ref{fig:6-31Gbarall}], the least important feature is the trace of the $\alpha\beta$-block of ${}^2\Delta$; this feature is also unimportant for the specific case of the ground spin state of C$_2$ [Fig.~\ref{fig:6-31Gbarall_C2_1.7}]. It can be shown that the trace of the full cumulant 2RDM is negative for a correlated system; it can also be shown that the trace of the $\alpha\beta$-block of ${}^2\Delta$ should be exactly zero. In practice, however, we obtain small non-zero values for Tr(${}^2\Delta_{\alpha\beta})$ that reflect the convergence criteria we use in the v2RDM optimization. We have intentionally included these small, non-physical quantities in our model as a means of testing the ability of SHAP analysis to identify irrelevant features. Indeed, the SHAP values clearly and correctly reflect that this feature does not impact the ML model, which suggests this sort of analysis could be used for more elaborate feature engineering. When retraining the model in the absence of the Tr(${}^2\Delta_{\alpha\beta})$ feature, we obtain incrementally improved results, with less overfitting.
This reduction in overfitting is highlighted by reduction of the absolute deviation between train and test RMSEs of 4.6253$\times 10^{-4}$ $E_{\text{h}}$ of the original model and 3.4936$\times 10^{-4}$ $E_{\text{h}}$ for the updated model (see Fig. \ref{fig:rm_tr_6-31G_bigstack_error}).

The SHAP-analysis-based trends we have highlighted for the 6-31G dataset are consistent with those observed in the STO-3G (Fig. \ref{fig:STO3Gbarall}) and cc-pVDZ (Fig. \ref{fig:ccpvdzbarall}) datasets, which are compared with the 6-31G data in Fig. \ref{fig:ALLSHAPCHANGE}.
As discussed above, the average violation of T2 is among the most important features in the STO-3G and cc-pVDZ models (actually, according the the SHAP analysis, it is the most important feature in these models), and the entropies of ${}^1{\bf Q}_{\alpha}$ and ${}^1{\bf Q}_{\beta}$ are also among the three most important features for the STO-3G dataset. For the cc-pVDZ dataset, only the entropy of ${}^1{\bf Q}_\beta$ appears among the three most important features. 
%For the STO-3G model, like the 6-31G data, the entropies of $\alpha$- and $\beta$-blocks of the one-hole RDM ($^{1}\mathbf{Q}$) also appear in the top three most important features.
%Unlike the STO-3G and 6-31G data, only the $\beta$- portions of $^{1}\mathbf{Q}$ appear in the top three most important features for the cc-pVDZ model.
As depicted in Fig.~\ref{fig:ALLSHAPCHANGE} we can also see that Tr(${}^2\Delta_{\alpha\beta})$ and violations of the T1 condition are, again, among the least important features in the datasets. 
Overall, the SHAP values appear to be useful for providing insights into the relationships between the physical information introduced through the DDv2RDM feature set and the target value.

